\documentclass[twocolumn,english,showpacs,aps,prl]{revtex4}
\usepackage{graphicx}

\makeatletter
\usepackage{color}
\usepackage{amssymb}
\usepackage{amsmath}
\usepackage{hyperref}
\hypersetup{
backref,
colorlinks=true,
citecolor=blue,
linkcolor=blue,
filecolor=blue,
urlcolor=blue}

\makeatother
\begin{document}

\title{Effects of dissipation on a quantum critical point with disorder}

\author{Jos\'e A. Hoyos}

\affiliation{Department of Physics, University of Missouri-Rolla, Rolla, MO 65409,
USA}

\author{Chetan Kotabage}

\affiliation{Department of Physics, University of Missouri-Rolla, Rolla, MO 65409,
USA}

\author{Thomas Vojta}

\affiliation{Department of Physics, University of Missouri-Rolla, Rolla, MO 65409,
USA}

\begin{abstract}
We study the effects of dissipation on a disordered quantum phase transition with O$(N)$
order parameter symmetry by applying a strong-disorder renormalization group to the
Landau-Ginzburg-Wilson field theory of the problem. We find that Ohmic dissipation
results in a non-perturbative infinite-randomness critical point with unconventional
activated dynamical scaling while superohmic damping leads to conventional behavior. We
discuss applications to the superconductor-metal transition in nanowires and to Hertz'
theory of the itinerant antiferromagnetic transition.
\end{abstract}


\date{12 October 2007}
\maketitle

The low-temperature properties of quantum many-particle systems are
often sensitive to small amounts of impurities or defects. Close to
quantum phase transitions (QPTs), the interplay between quantum fluctuations
and random fluctuations due to disorder can destabilize the conventional
critical behavior, leading to exotic phenomena such as quantum Griffiths
effects~\cite{thill-huse-physa95,rieger-young-prb96} and infinite-randomness
critical points~\cite{fisher92-95} as well as smeared phase transitions~\cite{vojta-prl03}
(for a recent review see, e.g., Ref.~\cite{vojta-review06}).

In particular, the QPTs in disordered quantum Ising magnets are governed by
infinite-randomness critical points~\cite{fisher92-95,motrunich-ising2d} which display
slow \emph{activated} dynamical scaling. In a dissipative environment, the dynamics
becomes even slower. In the experimentally relevant case of Ohmic dissipation, the
tunneling of sufficiently large droplets (the ones normally responsible for Griffiths
phenomena) is completely suppressed \cite{castroneto-jones-prb00,MMS-prl01-prb02}. As a
result, the sharp quantum phase transition is destroyed by smearing~\cite{vojta-prl03}.

In contrast, in dissipationless systems with \emph{continuous} O$(N)$ order parameter
symmetry, disorder does \emph{not} induce exotic infinite-randomness behavior in
dimensions $d>1$ ~\cite{insulator-AFM}. This changes in the presence of Ohmic
dissipation. It was recently shown that large locally ordered droplets are not frozen (in
contrast to the Ising case. Instead they display the exponentially slow dynamics associated with a
quantum Griffiths phase \cite{vojta-schmalian-prb05}. This leads to the important
question of whether the QPTs of
continuous symmetry order parameters with Ohmic dissipation are also of
infinite-randomness type.

\begin{figure}
\begin{center}\includegraphics[%
  clip,
  width=0.7\columnwidth,
  keepaspectratio]{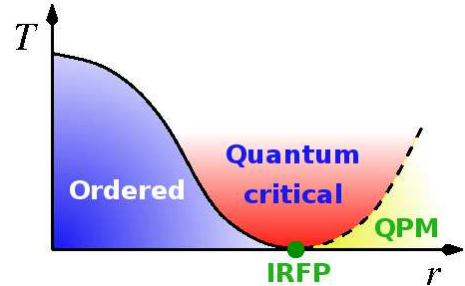}
\end{center}
\caption{(color online) Temperature--coupling phase diagram for Ohmic dissipation. IRFP
denotes the infinite-randomness critical point. The phase boundary (solid) and the
crossover line (dashed) between the quantum critical and quantum paramagnetic (QPM)
regions take unusual exponential forms leading to a wide quantum critical region. Both
phases contain Griffiths regions near the IRFP~\cite{vojta-schmalian-prb05}.
\label{cap:phase-diagram}}
\end{figure}

In this Letter, we answer this question and elucidate the nature of the transition by
applying a strong-disorder renormalization group (RG) to the  Landau-Ginzburg-Wilson
(LGW) order-parameter field theory of the problem. Our results are summarized as follows:
The QPT is controlled by an exotic infinite-randomness fixed point in the universality
class of the random transverse-field Ising model. The dynamical scaling is activated
rather than power-law, i.e., correlation time $\tau$ and correlation length $\xi$ are
related via $\ln\tau\sim\xi^{\psi}$, with $\psi$ the tunneling exponent. With decreasing
temperature, the order parameter susceptibility diverges as $\chi
\sim[\ln(1/T)]^{2\phi-d/\psi}/T$, and the specific heat vanishes as
$C\sim[\ln(1/T)]^{-d/\psi}$. Here, $\phi$ is the cluster size exponent. Close to the QPT,
the finite-temperature phase boundary takes the unusual form $T_{c}\sim\exp(-{\rm
const}\left|r\right|^{-\nu\psi})$ with $r$ the dimensionless distance from the QPT and
$\nu$ the correlation length exponent. The exponents $\psi$, $\phi$, and $\nu$ are
universal and identical to those of the random transverse-field Ising model. The
resulting phase diagram is shown in Fig.~\ref{cap:phase-diagram}.

Our starting point is a quantum LGW free energy functional
for an  $N$-component ($N>1$) order parameter $\varphi$ in $d$
dimensions. The clean action reads
\begin{equation}
S=\int{\rm d}y{\rm d}x ~\varphi(x)\Gamma(x,y)\varphi(y)+\frac{u}{2N}\int{\rm d}x~\varphi^{4}(x)\,,\label{eq:clean-action}
\end{equation}
 where $x\equiv(\mathbf{x},\tau)$ comprises imaginary time $\tau$
and position $\mathbf{x}$, $\int{\rm d}x\equiv\int{\rm d}\mathbf{x}\int_{0}^{1/T}{\rm d}\tau$,
and $\Gamma(x,y)$ is the bare inverse propagator (two-point vertex)
whose Fourier transform reads $\Gamma(\mathbf{q},\omega_{n})=r+\xi_{0}^{2}\mathbf{q}^{2}+\gamma\left|\omega_{n}\right|^{2/z_{0}}$
with $r$ the bare distance from criticality (the bare gap). $\xi_{0}$
is a microscopic length scale, and $\omega_{n}$ is a Matsubara frequency.
The damping coefficient $\gamma$ depends on the coupling of the order parameter to
the dissipative bath and the spectral density of the bath modes. We are
mostly interested in overdamped
(Ohmic) spin dynamics ($z_{0}=2$). However, to demonstrate the special
role of $z_{0}=2$, we also consider variable $z_{0}$. Quenched disorder
can be introduced by making the distance from criticality $r$ a random function
of position $r\rightarrow r+\delta r(\mathbf{x})$.
Analogously, disorder is introduced into $\xi_{0}$ and/or $\gamma$.

To apply the real-space based strong-disorder RG~\cite{MDH-prl-prb,igloi-review},
we discretize the action (\ref{eq:clean-action}) by defining discrete
coordinates $\mathbf{x}_{j}$ and rotor variables $\varphi_{j}(\tau)$.
It is important to note that the rotors do \emph{not} describe individual microscopic
degrees of freedom, they rather represent the average order parameter in a volume
$\Delta V$ large compared to $\xi_{0}$ but small compared to the
correlation length $\xi$, i.e., $\varphi_{j}(\tau)=\int_{\Delta V}{\rm d}\mathbf{y}\,\varphi(\mathbf{x}_{j}+\mathbf{y})$.

We first consider the large-$N$ limit of our action where all calculations
can be carried out explicitly. We will later show that the results
apply to all $N>1$. In the large-$N$ limit, the discrete action
reads \begin{eqnarray}
S & = & \frac{T}{E_{0}}\sum_{i}\sum_{\omega_{n}}\left(r_{i}+\lambda_{i}+\gamma_{i}\left|\omega_{n}\right|^{2/z_{0}}\right)\left|\phi_{i}(\omega_{n})\right|^{2}\nonumber \\
 &  & -\frac{T}{E_{0}}\sum_{\left\langle i,j\right\rangle }\sum_{\omega_{n}}\phi_{i}(-\omega_{n})J_{ij}\phi_{j}(\omega_{n})\,,\label{eq:action}\end{eqnarray}
 where $r_{i}$, $\gamma_{i}>0$ and the nearest-neighbor interactions
$J_{ij}>0$ are random quantities, $E_{0}$ is a microscopic energy
scale used to make the field dimensionless, and $\phi_{j}(\omega_{n})=E_{0}\int_{0}^{1/T}\varphi_{j}(\tau)e^{i\omega_{n}\tau}{\rm d}\tau$.
The Lagrange multiplier $\lambda_{i}$ enforces the large-$N$ constraint
$\langle(\varphi_{i}^{(k)}(\tau))^{2}\rangle=1$ for each order parameter
component $\varphi_{i}^{(k)}$. The local distance from criticality,
$\epsilon_{i}=r_{i}+\lambda_{i}$,
contains all single-site renormalizations and is thus always positive.

The basic idea of the strong-disorder (Ma-Dasgupta-Hu) RG is to successively
integrate out local high-energy degrees of freedom~\cite{MDH-prl-prb,fisher92-95,igloi-review}.
Here, the competing local energies are the local gaps $\epsilon_{i}$
and the interactions $J_{ij}$. In the bare theory (\ref{eq:action}),
they are independent random variables with distributions $Q(\epsilon)$
and $P(J)$, respectively.

In each RG step, we choose the largest local energy $\Omega=\max\{\epsilon_{i},J_{ij}\}$.
If it is a gap, say $\epsilon_{2}$, the unperturbed part of the action
is $S_{0}=(T/E_{0})\sum_{\omega_{n}}(\epsilon_{2}+\gamma_{2}\left|\omega_{n}\right|^{2/z_{0}})|\phi_{2}(\omega_{n})|^{2}$.
The coupling of $\phi_{2}$ to the neighboring sites, $S_{1}=-(T/E_{0})\sum_{j,\omega_{n}}J_{2j}\phi_{2}(-\omega_{n})\phi_{j}(\omega_{n})$,
is treated perturbatively. Keeping only the leading low-frequency
terms that arise in 2nd order of the cumulant expansion, we obtain
new interactions $\tilde{S}=-(T/E_{0})\sum_{\omega_{n}}\phi_{i}(-\omega_{n})\tilde{J}_{ij}\phi_{j}(\omega_{n})$
between all sites that used to couple to $\phi_{2}$, with \begin{equation}
\tilde{J}_{ij}=J_{ij}+\frac{J_{i2}J_{2j}}{\epsilon_{2}}\,.\label{eq:J-tilde}\end{equation}
 At the end of the RG step, $\phi_{2}$ is dropped from the action.

If the largest local energy is an interaction, say $J_{23}$, we solve the two-site
cluster
$S_{0}=(T/E_{0})\sum_{\omega_{n}}\sum_{i=2,3}(\epsilon_{i}+\gamma_{i}\left|\omega_{n}\right|^{2/z_{0}})|\phi_{i}(\omega_{n})|^{2}-(T/E_{0})\sum_{\omega_{n}}J_{23}\phi_{2}(-\omega_{n})\phi_{3}(\omega_{n})$
exactly while treating the interactions with all other sites as perturbations. The
calculation is straightforward but lengthy; details will be published elsewhere. For
$J_{23}\gg\epsilon_{2},\,\epsilon_{3}$, the two rotors $\phi_{2}$ and $\phi_{3}$ are
essentially parallel; thus they can be replaced by a single rotor $\tilde{\phi}_{2}$ with
effective renormalized action
$\tilde{S}=\left(T/E_{0}\right)\sum_{\omega_{n}}(\tilde{\epsilon_{2}}+\tilde{\gamma_{2}}\left|\omega_{n}\right|^{2/z_{0}})|\tilde{\phi_{2}}(\omega_{n})|^{2}$.
For Ohmic dissipation, $z_{0}=2$, the renormalized gap is given by
\begin{equation}
\tilde{\epsilon}_{2}=2\frac{\epsilon_{2}\epsilon_{3}}{J_{23}}\,,\label{eq:e-tilde}\end{equation}
 implying the relation $\tilde{\gamma_{2}}=\gamma_{2}+\gamma_{3}$
for the damping constants. The new rotor represents a cluster with
effective moment (number of sites represented) \begin{equation}
\tilde{\mu}_{2}=\mu_{2}+\mu_{3}\,.\label{eq:moment}\end{equation}
 The renormalized interactions of the new rotor with each of the remaining
ones are given by \begin{equation}
\tilde{J}_{2j}=J_{2j}+J_{3j}\,.\label{eq:J-tilde-cluster}\end{equation}

The net result of the RG step is the elimination of one site and the
reduction of the energy scale $\Omega$ together with renormalizations
and reconnections of the lattice. Since the structure of the RG recursion
relations (\ref{eq:J-tilde}-\ref{eq:J-tilde-cluster})
is identical to that of the random transverse-field Ising
model~\cite{fisher92-95,motrunich-ising2d,factor2},
we conclude that our system belongs to the same universality class.

In $d=1$, the RG step does not change the lattice topology.
One can thus derive flow equations for the individual probability
distributions of $\epsilon$ and $J$ and solve them analytically
~\cite{fisher92-95}. In $d>1$, new couplings are generated in each RG
step, and an analytical solution is impossible. However, by implementing
the recursion relations (\ref{eq:J-tilde}-\ref{eq:J-tilde-cluster}) numerically,
Motrunich et al.\ \cite{motrunich-ising2d} showed that there is a fixed
point in the full joint distribution of the $\epsilon$ and $J$ that
corresponds to the critical point of the system. In both cases, the
critical point is of infinite-randomness type. At criticality,
the distribution of the $\epsilon_{i}$ and $J_{ij}$ becomes singular
and broadens without limit as $\Omega\rightarrow0$
under renormalization which also provides an \emph{a posteriori} justification
for using the perturbative RG recursion relations (\ref{eq:J-tilde}-\ref{eq:J-tilde-cluster}).
One may be concerned about the initial stages of the RG in a weakly
disordered system, where the strong-disorder method is not very accurate.
However, perturbative RG studies~\cite{kirkpatrick-belitz-prl96,
NVBK-prl-prb99} showed that there is
\emph{no} stable weak-disorder fixed point; instead the perturbative RG shows runaway flow
towards large disorder.

We thus conclude that the infinite-randomness fixed point found here is \emph{universal}
and controls the transition for all nonzero disorder strength. Its critical behavior is
characterized by three exponents $\psi$, $\phi$, and $\nu$. The tunneling exponent $\psi$
controls the dynamical scaling, i.e., the relation between length scale $L$ and energy
scale $\Omega$, which is of activated rather than power-law type
\begin{equation}
\ln(1/\Omega)\sim L^{\psi}\,.\label{eq:activated}
\end{equation}
 It also controls the density $n_{\Omega}$ of surviving clusters
via $n_{\Omega}\sim[\ln(1/\Omega)]^{-d/\psi}$. $\phi$ describes
how the typical moment $\mu$ of a surviving cluster depends on $\Omega$,
\begin{equation}
\mu\sim\ln^{\phi}\left(1/\Omega\right)\,,\label{eq:moment-scaling}\end{equation}
 while $\nu$ determines how the correlation length $\xi$ depends
on the distance $r$ from criticality via $\xi\sim|r|^{-\nu}$. In one space dimension, the
exponents are known exactly from Fisher's analytical solution \cite{fisher92-95}:
$\psi=1/2$, $\phi=(1+\sqrt{5})/2$ and $\nu=2$. In two dimensions, they were determined
numerically \cite{motrunich-ising2d,lin-etal}, yielding $\psi=0.42\ldots0.6$,
$\phi=1.7\ldots2.5$ and $\nu=1.07\ldots1.25$. For $d=3$, the scaling towards an
infinite-randomness fixed point has been confirmed \cite{motrunich-ising2d}, but
estimates of the exponent values are still lacking.

We emphasize the particular role played by the Ohmic dissipation ($z_{0}=2$) of the
magnetic modes. To this end, we consider how the recursion relation (\ref{eq:e-tilde}) is
modified for $z_{0}\ne2$. For the superohmic case, $z_{0}<2$, we find
\begin{equation}
\tilde{\epsilon_{2}}^{-x}=\alpha\left[\epsilon_{2}^{-x}+\epsilon_{3}^{-x}\right]+\mathcal{O}\left(J_{23}^{-x}\right)\,,\label{eq:e-tilde-other-z0}\end{equation}
 where $x=(2-z_{0})/z_{0}$ and $\alpha$ is a constant~\cite{altman-brayali}.
Thus, the multiplicative structure of (\ref{eq:e-tilde}) is replaced
by an additive one. As a result, the local gaps $\epsilon$ are only
weakly renormalized for $z_{0}<2$. Near criticality, the distribution
of the interactions $J$ becomes extremely singular while that of
the gaps $\epsilon$ remains narrow. The critical point is therefore
not of infinite-randomness type but conventional with power-law scaling
$\tau\sim\xi^{z}$, although the dynamical exponent $z$ can become
arbitrarily large as $z_{0}\rightarrow2^{-}$. (Similar behavior was
found at a percolation QPT~\cite{vojta-schmalian-prl05}.)  For the subohmic case, $z_{0}>2$,
the sharp transition is destroyed by smearing because rare regions
can statically order independently from each other~\cite{vojta-review06,vojta-schmalian-prb05}.

We now turn to the behavior of observables which is similar to the
random transverse-field Ising model~\cite{fisher92-95,motrunich-ising2d}.
However, there are a few differences caused by the order parameter
symmetry and the damping of the modes. Summing over all surviving
clusters using (\ref{eq:activated}) and (\ref{eq:moment-scaling})
gives unusual scaling forms for the order parameter susceptibility
$\chi$ and the specific heat, \begin{align}
\chi(r,T) & =\frac{1}{T}\left[\ln(1/T)\right]^{2\phi-d/\psi}\Theta_{\chi}\left(r^{\nu\psi}\ln(1/T)\right)\,,\label{eq:chi}\\
C(r,T) & =\phantom{\frac{1}{T}}\left[\ln(1/T)\right]^{-d/\psi}\Theta_{C}\left(r^{\nu\psi}\ln(1/T)\right)\,,\end{align}
 where $\Theta_{\chi}$ and $\Theta_{C}$ are universal scaling functions.
At criticality, this leads to $C\sim[\ln(1/T)]^{-d/\psi}$ and $\chi\sim[\ln(1/T)]^{2\phi-d/\psi}/T$.
The dynamic order parameter susceptibility at criticality can be derived
similarly. On the real frequency axis, $\Im\chi(\omega+i0)=[\ln(1/\omega)]^{2\phi-d/\psi}/\omega$.
This implies that low-temperature inelastic scattering experiments
at the location of the order parameter Bragg peak should see a sharp upturn
in the scattering intensity $\sim[\ln(1/\omega)]^{2\phi-d/\psi}/\omega$
with $\omega\rightarrow0$.
The scaling form (\ref{eq:chi}) of the susceptibility can also be used to infer the shape
of the phase boundary close to the QPT. The finite-temperature transition corresponds to
a singularity in $\Theta_{\chi}(x)$ at some nonzero argument $x_{c}$. This yields the
unusual form $T_{c}\sim\exp(-{\rm const}\left|r\right|^{-\nu\psi})$ shown in Fig.\ 1. The
crossover line between the quantum critical and quantum paramagnetic regions displays
analogous behavior.

The infinite randomness at criticality leads to peculiar behavior of the correlation
functions. The \emph{average} correlation function $\overline{G}(\mathbf{x})$ is
dominated by the rare events of two distant sites belonging to the same surviving
cluster. This yields~\cite{fisher92-95,motrunich-ising2d}
$\overline{G}\left(\mathbf{x}\right)\sim|\mathbf{x}|^{-2\left(d-\phi\psi\right)}$. In
contrast, a typical pair of sites is not in the same cluster, and develops exponentially
weak correlations, $-\ln G_{{\rm typ}}\left(\mathbf{x}\right)\sim|\mathbf{x}|^{\psi}$.

Our explicit calculations are for the large-$N$ limit of the O$(N)$ LGW theory. To
discuss their relevance for finite $N$, we contrast the cases of Ising ($N=1$) and
continuous ($N>1$) symmetries. In the former, sufficiently strong Ohmic dissipation
freezes the magnetic droplets (the localization transition in the dissipative two-state
system~\cite{Leggett}) leading to a destruction of the sharp transition by
smearing~\cite{vojta-prl03}. Recently, this was confirmed in a numerical strong-disorder
RG~\cite{schehr-rieger}. In contrast, for $N>1$, isolated droplets continue to fluctuate
but with a tunneling rate (gap) that is exponentially small in their
size~\cite{vojta-schmalian-prb05} because O$(N)$ clusters are right \emph{at} the lower
critical dimension of the transition. This exponential size dependence of the gap
requires the multiplicative structure of the recursion (\ref{eq:e-tilde}) for the merging
of two rotors. We conclude that this multiplicative structure is valid for \emph{all}
$N>1$. Since (\ref{eq:J-tilde}) just reflects standard perturbation theory, all our RG
recursion relations and the resulting infinite-randomness critical point apply to the
general O$(N)$ case with $N>1$. This has been confirmed for undamped dynamics ($z_0=1$)
in the case of O$(2)$ symmetry and for a general O$(N)$ rotor model in Refs.\
\cite{altman-brayali}.

In the remaining paragraphs, we summarize our results, we discuss applications, and we
consider open questions. We have studied the effects of dissipation on the quantum phase
transition in a quenched disordered system with O$(N)$ symmetric order parameter. For
Ohmic dissipation, we have found an infinite-randomness fixed point while the behavior
for the superohmic case (including undamped dynamics) is conventional. For subohmic
dissipation, the quantum phase transition is destroyed by smearing. This must be
contrasted with the case of Ising symmetry for which undamped dynamics already leads to
an infinite randomness fixed point \cite{fisher92-95,motrunich-ising2d} while Ohmic
dissipation causes a smeared transition \cite{vojta-prl03}. All of these results are in
agreement with a general classification \cite{vojta-review06} of phase transitions in the
presence of weak disorder based on the effective dimensionality of rare regions: If
finite-size regions are exactly at the lower critical dimension, the critical point is of
infinite-randomness type. If they are below the lower critical dimension, the behavior is
conventional; and if they can order (freeze) independently, the transition is smeared.

Our theory has several applications. For instance, the superconductor-metal quantum phase
transition observed in thin nanowires \cite{M-SC-QPT-EXP} was studied using a LGW theory
analogous to (\ref{eq:clean-action}) in one dimension with an O$(2)$ (complex) order
parameter and Ohmic dissipation \cite{M-SC-QPT}. The effects of disorder on the
thermodynamics of this problem are described by our theory. Transport properties can also
be calculated using the methods of Refs.\ \cite{IRFP-transport}. Analogously, our theory
should apply to arrays of resistively shunted Josephson junctions.

A second potential application is the Hertz-Millis theory \cite{hertz-prb76,millis-prb93}
of the (incommensurate) itinerant antiferromagnetic transition. In this theory, a LGW
free energy analogous to (\ref{eq:clean-action}) is derived from a microscopic electron
Hamiltonian $H=H_{\rm band} + H_{\rm int} + H_{\rm dis}$ consisting of a nontrivial band
structure $H_{\rm band}$, a Hubbard-like interaction $H_{\rm int}$ and a random potential
$H_{\rm dis}$ by integrating out the fermionic degrees of freedom in favor of the bosonic
order-parameter field. The validity of the Hertz-Millis approach to these transitions is
still a controversial question as several experiments, in particular in heavy fermion
materials \cite{stewart-rmp01}, have shown marked differences from the predicted
behavior. Disorder effects are a much-discussed possible reason for these discrepancies
\cite{stewart-rmp01,miranda-dobrosavljevic-rpp05}. Our theory provides explicit results
on how dissipation and disorder can yield activated dynamics, quantum Griffiths
phenomena, and non-Fermi liquid behavior. This should make an experimental verification
or falsification of the disorder scenario much easier.
Note that attention must be paid to the long-range RKKY part of the interaction neglected
in (\ref{eq:clean-action}). It can produce an extra subohmic dissipation of locally
ordered clusters \cite{dobrosavljevic-miranda-prl05} which leads to freezing into a
``cluster glass'' phase~\cite{case-dobrosavljevic} at a low non-universal temperature
$T_{{\rm CG}}$ determined by the strength of the subleading RKKY interactions. However,
the infinite-randomness fixed point controls observables in the broad quantum critical
region above.

We thank E.~Miranda, J.~Schmalian and S.~Sachdev
for useful discussions. This work was supported by NSF under grant
no. DMR-0339147 and by Research Corporation. Parts of the research
have been performed at the Aspen Center for Physics.

\end{document}